\documentstyle[equations,fleqn]{mn}

\newif\ifpsfiles\psfilesfalse
\newlength{\colwidthf}
\newlength{\hcolwidthf}
\newlength{\colwidth}
\newlength{\hcolwidth}
\newcommand{\eq}[1]{
	\begin{equation}
	#1
	\end{equation}
}
\newcommand{\eqn}[2]{
	\begin{equation}
	#2
	\label{#1}
	\end{equation}
}

\def\spose#1{\hbox to 0pt{#1\hss}}
\def\lta{\mathrel{\spose{\lower 3pt\hbox{$\mathchar"218$}}
     \raise 2.0pt\hbox{$\mathchar"13C$}}}
\def\gta{\mathrel{\spose{\lower 3pt\hbox{$\mathchar"218$}}
     \raise 2.0pt\hbox{$\mathchar"13E$}}}
\def\equal{\! = \!}

\def\=#1{\overline{#1}}

\input epsf

\def\etal{et~al.\ }

\def\d{\rm d}
\def\D{\rm D}
\def\KPC{\rm kpc}
\def\PC{\rm pc}
\def\KM{\rm km}
\def\SEC{\rm s}
\def\KMS{\KM \SEC^{-1}}
\def\YR{\rm yr}
\def\yr{\YR}
\def\GYR{\rm Gyr}
\def\Gyr{\GYR}
\def\GYRS{\rm Gyrs}
\def\MSUN{M_\odot}

\def\msun{\MSUN}

\def\vek#1{{\bmath #1}}
\def\pot{\phi}
\def\div{{\rm \nabla\cdot\/}}
\def\rot{{\rm\bf \nabla\times\/}}

\def\csnd{c_{\rm s}}
\def\rhog{\Sigma}
\def\rhoginit{\Sigma_{0}}

\def\aver#1{\langle#1\rangle}

\def\hdivv{h\/\,\div\vek v}
\def\divv{\div \vek v}
\def\rotv{\rot \vek v}
\def\rhopre{\rho_{\rm pre}}
\def\rhopost{\rho_{\rm post}}
\def\vparas{v_{\parallel \rm s}}
\def\vperps{v_{\perp \rm s}}
\def\vparaspre{v_{\parallel \rm s, pre}}
\def\vparaspost{v_{\parallel \rm s, post}}

\def\runcmp{A-1} 	
\def\runbest{S-2}	

\def\runBB{S-1} 	
\def\runBD{S-2} 	
\def\runBA{S-3} 	
\def\runBE{S-4}		
\def\runBC{S-5} 	

\def\runBF{S-6} 	
\def\runBG{S-7} 	

\def\runAB{A-2} 	
\def\runAC{A-3} 	

\author[P.~Englmaier and O.E.~Gerhard]{Peter~Englmaier and Ortwin~Gerhard}

\title{Two modes of gas flow in a single barred galaxy}

\begin{document}
\maketitle
\begin{abstract}
\parindent=10pt

We investigate stationary gas flows in a fixed, rotating barred
potential.  The gas is assumed to be isothermal with an effective
sound speed $\csnd$, and the equations of motion are solved with
smoothed particle hydrodynamics (SPH).  Since the thermal energy in
cloud random motions is negligible compared to the orbital kinetic
energy, no dependence of the flow on $\csnd$ is expected. However, this
is not the case when shocks are involved.

For low values of $\csnd$ an open, off-axis shock flow forms that is
characteristic for potentials with an inner Lindblad resonance (ILR).
Through this shock the gas streams inwards from $x_1$ to $x_2$-orbits.
At high sound speeds the gas arranges itself in a different, on-axis
shock flow pattern. In this case, there is no gas on $x_2$-orbits,
demonstrating that the gas can behave as if there were no ILR. The
critical effective sound speed dividing the two regimes is in the
range of values observed in the Milky Way. 

We give a heuristic explanation for this effect. A possible
consequence is that star formation may change the structure of
the flow by which it was initiated. Low-mass galaxies should
predominantly be in the on-axis regime.

A brief comparison of our SPH results with those from a grid-based 
hydrodynamic code is also given.
\end{abstract}
\begin{keywords}
Galaxies: spiral --- Galaxies: interstellar matter ---
Interstellar medium: kinematics and dynamics --- Hydrodynamics
--- Shock waves 
\end{keywords}
\section{Introduction}

Gas kinematic observations in the best-studied barred galaxies imply
highly non-circular gas motions. Some examples are NGC 5383 (Duval \&
Athanassoula 1983), NGC 7496 (Pence \& Blackman 1984b), NGC 1365
(Teuben \etal 1986), and our Galaxy (Liszt \& Burton 1980, Binney
\etal 1991).  These non-circular motions are believed to be associated
with shocks; large velocity gradients in support of this
interpretation have been observed in NGC 6221 (Pence \& Blackman
1984a), NGC 1365 (Lindblad \& J\"ors\"ater 1987) and NGC 3095 (Weiner,
Williams \& Sellwood 1993). The shocks can be traced on optical images
by the dust lanes that mark swept-up dense gas at the leading edges of
the bar (Pence \& Blackman 1984a, Athanassoula 1992). In IC 342,
millimeter observations of the molecular gas distribution and velocity
field suggest that the gas in the shock ridges is falling into the
nucleus (Ishizuki \etal 1990).  Some estimates for the resulting mass
infall rates are $\sim 0.01-0.1\msun/\yr$ in our Galaxy (Gerhard \&
Binney 1993) and $\sim 4 \msun/\yr$ in NGC 7479 (Quillen \etal
1995). The infalling gas may settle on an inner ring and fuel
starburst activity (Ishizuki \etal 1990, Kenney \etal 1992).

Because the effective temperature (cloud velocity dispersion) of the
gas is much smaller than any orbital velocities, it is generally a
good approximation to think of quasi-stationary galactic gas flows in
terms of the periodic orbits in the underlying gravitational
potential, so long as these periodic orbits do not cross. Many
hydrodynamic simulations using a variety of numerical techniques have
confirmed this (Sander \& Huntley 1976; Schwarz 1981, 1984; Combes \&
Gerin 1985, Habe \& Ikeuchi 1985; van Albada 1985; Mulder \& Liem
1986; Athanassoula 1992; Friedli \& Benz 1993; Jenkins \& Binney
1994). When the periodic orbits do cross or intersect, then pressure
or viscous forces must always become important and ensure that the gas
streamlines match together to form a well-defined flow pattern. This
typically happens near the orbital resonances of the system, and it is
here that shocks are found to form.

The geometrical structure, the extent, and the strength of these
shocks depend on a variety of external parameters. Most important is
whether the gravitational potential has an Inner Lindblad Resonance
(ILR) and hence a family of $x_2$-orbits , but also the strength and
axis ratio of the bar are significant parameters (Roberts, van Albada
\& Huntley 1979, Sanders \& Tubbs 1980, Athanassoula 1992).

We have begun a study aimed at better understanding these gas flows,
and using them as tracers for the dynamical properties of our Galaxy
and barred galaxies in general. We have used a two-dimensional
smoothed particle hydrodynamics (SPH) method, based on the SPH-code of
Steinmetz \& M\"uller (1993) which Matthias Steinmetz kindly made
available. The extent to which the interstellar medium (ISM) in
galaxies can be modelled by simple gas dynamics has been discussed
recently by Sellwood \& Wilkinson (1993) and Binney \& Gerhard (1993).
In the present study, we have omitted the self-gravity of the fluid,
treating it as a tracer; the restriction to two dimensions is in order
to increase the resolution. Later, the SPH method will allow us to
study the effect of the gas' self-gravity, particularly in the
high-density regions that generally form in these flows.

In the course of this work, we have found that the shock properties
in barred potentials also depend on internal parameters used to model
the gas; specifically, on the sound speed if the gas is assumed to be
an isothermal fluid. This paper reports on these results. We first
describe the mass model, initial conditions, etc.\ (Section 2) and
give some details on the numerical method (Section 3). In Section 4 we
show that both off-axis and on-axis shocks may form in a
potential with an ILR, study the influence of some relevant
parameters, and briefly compare our results with those from grid-based
simulations.  Finally, in Section 5 we discuss some implications of
our main result.

\section{Description of the Models}

In this paper we consider gas flows in galaxy models with a given
gravitational potential for the stellar component, as detailed in
Section 2.1. Our assumptions concerning the initial conditions, the
hydrodynamics of the gas, and a simple gas recycling law for mimicking
star formation are given in Sections 2.2-2.4, while the numerical
method (Smooth Particle Hydrodynamics, SPH) is briefly described in
Section~3.

\subsection{Mass models for the stellar component}

In order to be able to compare our results from SPH with those from a
hydrodynamic grid method, we have adopted the analytic family of
models from Athanassoula (1992).  These models include a bulge, disk,
and bar component.  The bulge follows a Hubble profile
\eqn{rhobA}{
     \rho(r)=\rho_b(1+r^2/r_b^2)^{-3/2} 
}
and the stellar disk is a simple
Kuzmin--Toomre disk with surface density
\eqn{sigdA}{
   \Sigma_d(R)={R_d M_d\over 2 \pi} (R^2+R_d^2)^{-3/2}. 
}
The parameters are chosen such as to give an approximately flat
rotation curve:
 $R_d = 14.1\;\KPC$, $M_d=2.3\times
10^{11}\MSUN$, $r_b=0.33\;\KPC$, and $\rho_b=23.6\;\MSUN/\PC^3$.

The bar component is modelled as a simple Ferrers ellipsoid with density
\eqn{rhobarA}{
\rho(\vek X)=\left\{
       \begin{array}{ll}
      \rho_F (1-g^2) & \mbox{for } g^2<1\\
	0            & \mbox{elsewhere,}
       \end{array}
     \right.
}
 with $g^2\equal X^2/a_F^2+(Y^2+Z^2)/b_F^2$ and parameters $a_F\equal
5\;\KPC$, $b_F\equal 2\;\KPC$. We have considered two values for the
bar mass: one is specified by setting $\rho_F\equal
0.4476\;\MSUN/\PC^3$ (Model $A$, corresponding to Athanassoula's model
001), the other has one half that mass (our standard Model $S$).  To
place the L1-Lagrange point at $r_{\rm L1}=6.0\;\KPC$ we chose a
pattern speed of $\Omega_p\equal 35.3\;\KM\;\SEC^{-1}\;\KPC^{-1}$ for
the Model $A$ bar and $\Omega_p\equal32.0\;\KM\;\SEC^{-1}\;\KPC^{-1}$
for the Model $S$ bar. This corresponds to bar rotation periods of
$0.174\;\GYR$ and $0.192\;\GYR$, respectively.  Then from the $m\equal
0$--component of the potential, the corotation radius is $R_{\rm
CR}=5.8\;\KPC$ in both cases.

The rotation curves for both models and the contributions of
the individual components (from the respective $m\equal 0$--components
of the gravitational force in the equatorial plane) are shown in
Fig.~\ref{rotcurve}.

\begin{figure}
\epsfxsize=\colwidthf\ifpsfiles\epsfbox{fig1.ps}\fi
\caption[-]{Circular rotation curves in the equatorial plane:
total (full and dotted lines) for Models $A$ and $S$, and
contributions from the bulge (long-dashed), Ferrer's bar (short-dashed
for Model $A$, dotted for Model $S$), and disk (dash-dotted).}
\label{rotcurve}
\end{figure}

\begin{table}\label{tbl-models}
\begin{tabular}{l|l|l|l|l|l}
Model & $\csnd [\KM/\SEC]$ & Gas recycl. & $\zeta$ & $t [\Gyr]$\\
\hline\\
\runBB & 10 & no  & 1.2 & 0.6\\
\runBD & 15 & no  & 1.2 & 0.6\\
\runBA & 20 & no  & 1.2 & 0.6\\
\runBE & 25 & no  & 1.2 & 0.6\\
\runBC & 30 & no  & 1.2 & 0.6\\
\runBF & 15 & yes & 1.2 & 0.6\\
\runBG & 20 & yes & 1.2 & 0.6\\
\runcmp & 10 & no & 1.8 & 0.4\\
\runAB & 15 & yes & 1.2 & 0.6\\
\runAC & 20 & yes & 1.2 & 0.6\\
\hline
\end{tabular}
\caption[-]
{Parameters for the models computed.  For all runs we used $N\approx
2\;10^4$ particles in 2D and imposed point symmetry, giving a mean
resolution lengths of $h\lta 0.1\;\KPC$.  All simulations are
rotating clockwise. Table columns give (1) Model designation,
(2) effective sound speed of the gas, (3) whether gas recycling was
used, (4) constant determining the individual particle sizes, cf.~Section
3, and (5) time since start of the calculation for which
results are reported.}
\end{table}

\subsection{Initial conditions and symmetry}
In the present two-dimensional calculations the gas is initially
set up on circular orbits with constant surface density
$\rhoginit=1\;\MSUN\;\PC^{-2}$.  Outside some radius $r_{\rm cut}$ beyond
corotation the density is set to a constant, i.e., the pressure
gradients are set to zero.

The non-axisymmetric part of the potential is turned on gradually
within a time $\tau_{\rm on}$ equal to approximately one half the rotation
period of the bar. After continuing the calculation for some time $t$
given in Table 1 the gas flow is nearly stationary in the rotating
frame.

In these calculations we have assumed point symmetry with respect
to the origin. This effectively doubles the number of particles,
and increases the resolution by a factor $\sqrt{2}$.
We have checked that the results in simulations without the
assumed point symmetry are essentially identical.

\subsection{Gasdynamics}

For the gas we have assumed an isothermal equation of state with an
effective sound speed $\csnd$, representing the random motions in the
interstellar cloud medium. This approach is consistent with the ISM
model of Cowie (1980), who argued that the cloud fluid can be treated
like an isothermal gas if the clouds have an equilibrium mass
spectrum. The equlibrium is assumed to be maintained by a steady
supply of small clouds by supernovae. Using SPH as described in the
next section we are thus solving the Euler equations
\eq{
   {\D \vek{v}\over\D t}=
     {\partial\vek{v}\over\partial t}+(\vek{v}\cdot\vek{\nabla})\vek{v} =
   - \csnd^2 {\vek{\nabla} \Sigma\over\Sigma} - {\vek{\nabla}\pot}
}
for effective sound speeds of $\csnd=10-30\KMS$.

\subsection{Gas recycling}

Because the gas loses angular momentum to the stellar bar by
gravitational torques, a substantial accretion rate towards the centre
develops.  In run \runBB, for example, at a radius of $1\;\KPC$, the
accretion rate is approximately $7.6\times10^6\,\MSUN/\GYR$, and in
\runBC, it is $\sim 1.9 \times 10^7\,\MSUN/\GYR$, rising continuously
with sound speed between these two cases. For comparison, for
our initial surface density of $1 \MSUN/\PC^2$ the total gas mass
inside corotation is $1.06\times10^8\;\MSUN$.  Thus after $\sim
5-15\,\GYRS$ all of the gas would have been accreted to the centre. To
avoid this, we have in some of our models introduced a gas recycling
law which takes away gas in high density regions and adds material
uniformly over the disk:
\eqn{recycle}{
	\d \rhog/ \d t = \alpha \rhoginit^2 - \alpha \rhog^2.
}
This may crudely be thought of as simulating star formation in
dense regions and mass loss from preexisting bulge and disk stars. 
The star formation rate $\alpha$ is taken to be 
$0.3 \MSUN^{-1} \PC^2 \GYR^{-1}$. Equation~(\ref{recycle}) includes
a quadratic Schmidt (1959) law and a constant source term, but
the functional form was mainly chosen to be able to compare with
the grid-based simulations of Athanassoula (1992).

\section{Numerical method}

We have used a two-dimensional SPH method to solve the hydrodynamical
equations for the gas flow in the galaxy equatorial plane. Restricting
the calculation to two dimensions increases greatly the resolution we
can achieve, and this is important in the present problem for
resolving the shocks that dominate the evolution. Since we are
interested in isothermal shocks, we are assuming implicitly that the
gas can cool rapidly. We then expect that the gas can expand
vertically with speeds only of the order of the sound speed. Since
this is much smaller than the dynamical velocities, vertical motions
cannot affect the shock front even if this has finite width (such as
one or two cloud mean free paths). The gas will also only marginally
be able to expand out of a three-dimensional layer within a dynamical
time; hence the expansion of the post-shock gas can not lead it too
far above the downstream material that it will meet next.  Thus we
expect the two-dimensional description to be a reasonable
approximation. Notice that it is not possible to test this rigourously
with three-dimensional simulations of this particle number, unless
anisotropic particles are used; for to resolve the disk vertically,
one needs to make it substantially thicker than in reality, which
prolongs the vertical escape times.  We have done some such
three-dimensional simulations; in these the observed gas flows were
indeed similar to the two-dimensional ones.

As there are many variants of SPH (see the reviews of Benz 1990,
Monaghan 1992), we give a short overview of our code, which is based
on the code of Steinmetz\&M\"uller (1993). A few details on how the
method works will be useful for interpreting the results reported
below.  SPH solves the equations of motion by a Monte Carlo
integration. The integration points are not drawn by random, but are
the positions of the particles, as they happen to be. Ideally the
particles should be distributed in a glass-like structure to minimize
numerical errors.  Apart from this the particles have no physical
meaning.  The forces on each parcel of fluid can be calculated by
smearing out the properties of the particles over a few mean particle
distances.

For constant smoothing length $h$, the smoothing is done by folding
each field quantity $A$ such as temperature or pressure
with the kernel function $W(\vek{r})$
\eqn{sphmean}{
  \begin{eqalign}
	\aver{A(\vek{r})} &= \int{A(\vek{r}') W(\vek{r}' - \vek{r}, h) \d^2x'}
       + O(h^2) \\
	&\approx \sum_i{{m_i\over\Sigma_i} A_i W(\vek{r}_i-\vek{r},h)}.
  \end{eqalign}
}
Here $A_i$ and $\Sigma_i$ are evaluated at the position of particle $i$.
Thus the surface density is approximated by
\eqn{surfdens}{
   \aver{\Sigma(\vek{r})} = \sum_i{m_i W(\vek{r}_i-\vek{r},h)}. 
}
Only structure on scales larger than a few resolution lengths $h$
is meaningful.
We used the spline kernel
\eqn{splinekernel}{
	W(\vek{r} - \vek{r}',h)={W_0\over h^2} \left\{\begin{array}{ll}
	1 - {3\over2}u^2+{3\over4}u^3 & 0 \le u \le 1 \\
	{1\over4}(2-u)^3 & 1 < u \le 2 \\
	0 & u > 2 
	\end{array} \right.
}
with $u=|\vek{r} - \vek{r}'|/h$, and the constant $W_0$ is determined
from the normalization condition
\eq{
	1 = \int{W(\vek{r},h) \d^2x}.
}
The kernel describes the surface density distribution of one particle.

To increase resolution in high-density regions the particles are
assigned individual smoothing lengths
\eq{
 	h_i = \zeta \sqrt{m_i / \Sigma_i}
}
with $\zeta=1.2$ in most cases (see Table 1). In this case, the field
quantities at particle positions $\vek{r_i}$ are evaluated with the
symmetrized Kernel
\eq{
     W(\vek{r_i}-\vek{r_j},h_i,h_j) = \textstyle{1\over 2} \left\{
     W(\vek{r_i}-\vek{r_j},h_i) + W(\vek{r_i}-\vek{r_j},h_j) \right\}
}
to ensure momentum conservation. This symmetrized Kernel is also used
in the calculation of the surface density $\Sigma_i =
\aver{\Sigma(\vek{r_i})}$ at particle position $\vek{r_i}$. To evaluate
the field quantity at an arbitrary position $r$ the original 
Kernel~(\ref{splinekernel}) is used.

We have used the following simple approximation for adjusting the
particle smoothing length:
\eq{
   h^{(n+1)}_i=\sqrt{{N_{0}\over N^{(n)}_i}} h^{(n)}_i.
}
Here $N_{0}=4\pi\zeta^2$ is the desired number of neighbours 
within $2h_i$, and
$N^{(n)}_i$ the recorded number of neighbours of particle $i$ in the
$n$-th time step. This ensures that the number of neighbours of a
particle is approximately independent of time and of its location. In
our case with strong shocks the original method used by Steinmetz
(1993) was found to result in a larger number of neighbours in the
shock region and a loss in resolution.  In some cases we also observed
large oscillations in the particle size from time step to time step
when a particle entered the shock.

The Euler equations in SPH read
\begin{eqnarray}
   {\d \vek{v}_i\over \d t}\!\!\!\! &=& \!\!\!\!
                - \sum_j{m_j\left({P_i\over\Sigma_i^2}+{P_j\over
                \Sigma_j^2}+Q_{ij}\right)\nabla_i 
                  W(\vek{r}_i - {\vek{r}_j},h_i,h_j)} \nonumber\\
	& & \!\!\!\! -{\vek{\nabla}\pot_i}. \yesnumber
\end{eqnarray}
where $Q_{ij}$ is an artificial viscosity tensor (Monaghan \& Gingold 1983)
\eq{
	Q_{ij}=\left\{\begin{array}{ll}\displaystyle
		{-\alpha \csnd \mu_{ij} + \beta \mu_{ij}^2\over \Sigma_{ij}},
		& (\vek{r}_i-\vek{r}_j)\cdot(\vek{v}_i-\vek{v}_j)\le0\\
		0, &\mbox{otherwise,} 
	\end{array}
	\right.
}
\eq{
	\mu_{ij}={h_{ij} (\vek{r}_i-\vek{r}_j)\cdot(\vek{v}_i-\vek{v}_j)
	\over (\vek{r}_i-\vek{r}_j)^2 + \eta^2 },
}
with $\alpha\equal 1$, $\beta\equal 2$, $\eta\equal 0.1\;h_{ij}$,
$h_{ij}\equal (h_i+h_j)/2$, and $\Sigma_{ij}\equal(\Sigma_{i} +
\Sigma_{j})/2$.  The quantity $\mu_{ij}$ is an approximation for the
contribution of the particle pair $(i,j)$ to the local volume
compression $\hdivv$ over one resolution length $h$.

The artificial viscosity actually consists of two contributions.  The
term with the constant $\alpha$ corresponds to the bulk viscosity
whereas the term with the constant $\beta$ corresponds to the
von~Neumann-Richtmyer viscosity.  Both correspond to a `viscous
pressure' in the Euler equation, namely $\alpha \Sigma \csnd |\hdivv|$ 
and $\beta \Sigma (\hdivv)^2$ (Benz 1990).  The
von~Neumann-Richtmyer viscosity is necessary to simulate strong shocks
because it guarantees causality by transporting information with
signal speed $\sqrt{\beta}|\hdivv|$ across the shock fronts
(e.g., the flow is supersonically compressed ($\csnd < -\hdivv$).  
The bulk viscosity is used to damp post-shock oscillations.

As reported by Hernquist \& Katz (1989), the artificial viscosity tensor
does not vanish in pure shear flows. Therefore, some authors use the
modified artificial viscosity tensor proposed by Balsara (1995)
\eq{
	\tilde Q_{ij}=Q_{ij} {1\over2}(f_i + f_j)
}
with 
\eq{
	f_i = {|\aver{\divv}_i| \over |\aver{\divv}_i|
	   + |\aver{\rotv}_i| + 0.0001 \csnd/h_i }.
}
The effect of this is to reduce the viscosity in regions with large
values of $|\aver{\rotv}_i|$.  However, we have not found this
viscosity switch useful in the presence of shocks with strong
shear. In a sheared shock, both the divergence and the rotation of the
velocity field become large, thus the $f_i$ decrease below unity and
the viscosity becomes too small to treat the shocks adequately.

The gas recycling law (\ref{recycle}) is realized by adjusting the mass of
each particle $m_i$ at its individual timestep $n+1$ according to
\eq{
\begin{eqalign}
	m_i^{(n+1)} &= m_i^{(n)} 
        + \d t_i {\d m\over\d\Sigma}{\d\Sigma\over\d t}, \\
	&= m_i^{(n)}  
	+ \d t_{i} \left({h_i\over\zeta}\right)^2 
	\alpha(\rhoginit^2-{\rhog}_i^2).
\end{eqalign}
}
The particle velocities are kept constant.

\section{Results}

\begin{figure}
\epsfxsize=\colwidth \ifpsfiles\epsfbox{fig2.ps}\fi
\caption[-]{%
(a) Particle distribution in Model \runbest, for time 
$t=0.6\,\GYR$. The outline of the bar is given by the ellipse. 
(b) The velocity field for the same model. 
(c) Some $x_1$ and $x_2$ orbits useful for the interpretation of the 
gas flow. 
}
\label{flow-S}
\end{figure}

\subsection{Off-axis shocks at low sound speeds}

Fig.~\ref{flow-S} shows the distribution of gas particles, the velocity
field and the underlying orbital structure in Model \runbest.  This
model describes an isothermal fluid with a sound speed $\csnd=15\KMS$. The
velocity field was calculated on a regular grid, using the SPH
smoothing algorithm, and is shown in a rotating frame in which the bar
and the gas flow pattern are approximately stationary.

As is well-known, the flow is influenced strongly by the families of
periodic orbits in the underlying potential and the transitions
between them. The most important families here are the
$x_1$-orbits elongated along the bar and the central $x_2$-orbits
oriented perpendicular to the bar.  The outermost orbits shown in
Fig.~\ref{flow-S}c belong to a $4/1$-resonant family. See Contopoulos \&
Papayannopoulos (1980) for this notation.

The velocity field shows that the gas occupies $x_1$ orbits in the
main bar region and then shifts gradually to $x_2$ orbits in the inner
parts.  In this process it forms a shock along the leading edges of
the bar; in model \runbest\ the shock starts at $\approx 3.7\,\KPC$ and
continues inward with a small inclination angle with respect to nearby
$x_1$-orbits. From a particle point of view, the gas particles moving
through the shock exchange momentum with particles downstream to avoid
penetration of each other.

The gas moves inwards alongside the shock and then swings around the
disk of $x_2$-orbits towards the symmetric far-side branch of the
shock. The flow follows the $x_2$-orbits some way around the
centre, reaches a density maximum and opens like a spray as some
particles go into the disk, whereas others move outward again into the
far-side shock. With lower orbital energy, they then begin another
half-revolution. The gas thus loses further orbital energy and spirals
inwards in a few revolutions until it joins the gas on the outer
occupied $x_2$-orbits.  The resulting central ring of dense gas is
often called the $x_2$-disk.

Gas coming into the bar region from outside encounters a trailing
spiral shock wave which occurs at the intersection of colliding
streams on the outer $x_1$ orbits and the $4/1$ orbit family.  In
Model \runbest\ this shock front starts on the major axis at $\approx
3.7\,\KPC$ where the $x_1$ family meets the $4/1$ family. For definiteness,
we call the two types of shock the `cusped-orbit shock' and the 
`$4/1$ spiral shock'.

Fig.~\ref{shock1-S} shows the density and the velocities of particles
within one smoothing length $h$ of `Slit 1' plotted in
Fig.~\ref{flow-S}. The left side of Fig.~\ref{shock1-S} with small $s$
corresponds to the low density, pre-shock region inside the cusped
orbit. The top panel shows the gas surface density through the cusped
orbit shock along the slit. The points give the surface density
$\Sigma_i$ evaluated at the positions of individual particles; the
solid line shows the smoothed SPH value along the slit computed by
integrating over all overlapping particles [cf.~eq.~(\ref{surfdens})].
As determined from the latter, the density jump across the shock is
about a factor of two; because of resolution effects this is a lower
limit. The rise of the density from its pre-shock value occurs on a
length scale comparable to the smoothing length $h$; here $h$ is
approximately $140\PC$ upstream, $70\PC$ at the density peak and
$90\PC$ downstream.
The post-shock
region (near the peak of the density) appears still not resolved;
most of the right side of the diagram shows densities on streamlines
which are quite unrelated to the shock (see Fig.~\ref{flow-S}).

The lower two panels of Fig.~\ref{shock1-S} show the gas velocities
$v_{\parallel s}$ and $v_{\perp s}$ parallel and perpendicular to the
`slit' direction in Fig.~\ref{flow-S}a. At the shock surface
$\vparas$ is approximately the velocity with which a
supersonic collision occurs, $\vperps$ is approximately the
transverse shear component of the streaming velocity. It is seen that
$\vparas$ jumps by $\sim 100\KMS$ over the same length-scale
over which the density jump occurs. The full symbols in
Fig.~\ref{shock1-S} denote those particle positions at which $-\hdivv > \csnd$
 --- the local criterium for a shock to take place. For a
simple non-shearing, isothermal shock the Rankine-Hugoniot jump
conditions predict $\vparaspre\times\vparaspost \equal \csnd^2$; 
hence for $\Delta \vparas \simeq 100\KMS$ and $\csnd = 15 \KMS$ one finds 
$\vparaspre \simeq 102 \KMS$ and $\vparaspost \simeq 2 \KMS$ 
for the upstream pre-shock and downstream post-shock
components, which appears approximately correct. The corresponding
density jump is a factor of 50 and must then be clearly unresolved.

However, from the lower panel and also from Fig.~\ref{flow-S}b we see
that the gas flow in this region has a strong shear component and there
is likely also a discontinuity in the transverse $\vperps$ velocity across
the shock. Quantitatively, $\Delta \vperps \simeq 180\KMS$ across the
unresolved shock width. In this case, the jump conditions are not as
simple (Syer \& Narayan 1993).

\begin{figure}
\epsfxsize=\colwidth \ifpsfiles\epsfbox{fig3.ps}\fi
\caption[-]{%
(a) Surface density, (b) velocity $\vparas$ along, and (c)
velocity $\vperps$ perpendicular to `Slit 1' in Fig.~\ref{flow-S},
for all particles within one smoothing length $h$ of the slit. This
slit intersects the cusped orbit shock approximately perpendicularly.
The diamonds give the surface density $\Sigma_i$ evaluated at the 
positions of individual particles; the
solid line shows the smoothed SPH value along the slit computed by
integrating over all overlapping particles.
Full symbols denote particles which are deccelerated in the shock,
i.e., are compressed supersonically: $-\hdivv > \csnd$.
}
\label{shock1-S}
\end{figure}
\begin{figure}
\epsfxsize=\colwidth \ifpsfiles\epsfbox{fig4.ps}\fi
\caption[-]{%
Same as Fig.~\ref{shock1-S}, but for 'Slit 2' through the
$4/1$-spiral shock in Fig.~\ref{flow-S}.
}
\label{shock2-S}
\end{figure}

Fig.~\ref{shock2-S} shows analogous information for the $4/1$-spiral
shock. This shock is less complicated because shear velocities are
seen only in the post-shock region, but are essentially absent in the
shock region itself. The density goes up by a factor of four, and the
Mach number $\vparas/\csnd$ also jumps by a factor of four.  From the
Rankine-Hugoniot conditions we would expect
$\rhopre/\rhopost = 1/M^2$, so the maximum density
is again unresolved.

\subsection{On-axis shocks at high sound speeds}

So far we have adopted a sound speed of $\csnd = 15\;\KM/\SEC$ which is
consistent with observations of the turbulent speed of clouds in the
ISM. However, the observed values in the Galactic disk do vary, so it
is interesting to see whether the stationary flow pattern depends on
$\csnd$ or not.

From a theoretical point of view we might expect several regimes to
occur between two extreme cases: In the limit of negligible sound
speed every slight compression of the gas causes a shock and an
isothermal fluid would become effectively incompressible.  In the
other limit of high sound speeds no shocks can form at all.

Indeed, we found that there are two qualitatively different
quasi-stationary gas flow solutions in the potentials investigated.
The transition between these two configurations is shown in
Fig.~\ref{cs-seq}. At low sound speeds, the gas flow is as described
in the last section. At higher sound speeds ($\csnd\simeq 25-
30\;\KM\;\SEC^{-1}$ in our standard potential with circular velocity
$\approx200\KMS$) the $x_2$-orbits are no longer occupied and a
cusped orbit shock cannot form; instead, it is replaced by a
broken shock configuration near the long axis of the bar. This
structure resembles more the shocks that Athanassoula (1992) found to
occur in barred potentials with no Inner Lindblad Resonance (ILR),
while of course the potential used still has an ILR.  The spiral wave
also is no longer strong enough to form a 4/1-spiral shock.

A heuristic explanation for these results is as follows. Consider a
sequence of models such as in Fig.~\ref{cs-seq}, starting with a case
with well-developed off-axis shocks.  The speed with which the spray
of gas inside the cusped orbit meets the cusped orbit shock decreases
outwards (Fig.~2b). Thus the shock front should end where the velocity
$v_{\parallel s}$ in the shock frame becomes smaller than the sound speed. When
the sound speed is increased while keeping the potential constant,
the streaming velocities will to first order remain unchanged, so that
the shock front must move inwards. This means that it must move
towards $x_1$-orbits deeper in the potential well and closer to the
major axis. The second aspect is the gas inflow along the shock
which feeds gas onto the $x_2$-ring.  As the shock front moves closer
to the bar major axis, the inflowing gas will eventually no longer be
able to settle on $x_2$-orbits. Once $x_1$-orbits cannot be occupied, one
of the principal factors instrumental in building up the off-axis
shock has disappeared.

From Fig.~\ref{cs-seq}, we conclude that the transition between both
solutions in terms of the variation of $\csnd$ is more or less
continuous. E.~Athanassoula (private communication) has also found in
her grid simulations the existence of two distinct gas flow patterns;
however, she appears to see a very sharp transition at a critical
sound speed of $\csnd=(16.4\pm0.1)\;\KM\;\SEC^{-1}$ (in the potential
model A). It is not clear at present precisely which aspects of the
numerical schemes are responsible for this difference; nor whether any
of these schemes describes the dynamics of the ISM accurately enough
to say how the real ISM behaves in this respect.

An interesting aspect of the flow patterns at high sound speed is that
the gas inflow still forms a central disk; however this now consists
of gas on $x_1$-orbits rather than $x_2$-orbits.  This is only
possible because in our Ferrers bar potential the $x_1$ orbits are
not self-intersecting. If they did self-intersect, a still 
different flow pattern connected with strong accretion to the centre
would likely result.

%
\begin{figure*}
\epsfxsize=0.91\textwidth\ifpsfiles\epsfbox{fig5.ps}\fi
\caption[-]{%
A sequence of models with increasing sound speed as indicated in the
upper left corner of each panel (models \runBB\ to \runBC\ in Table
1).  Between $20$ and $25\;\KM/\SEC$ the flow pattern changes as
described in the text. The orbital structure in all the simulations is
the same; some closed orbits are shown in the lower right panel.
}
\label{cs-seq}
\end{figure*}
%

\subsection{Influence of initial situation}

\begin{figure}
\epsfxsize=0.95\colwidth\ifpsfiles\epsfbox{fig6.ps}\fi
\caption[-]{%
An on-axis shock solution switches to an off-axis shock flow 
when the sound speed is changed by hand and the massive central
$x_1$-disk is removed. Times are specified in $\GYR$ and the
bar rotation period is $0.174\GYR$.
}
\label{trans}
\end{figure}

We now ask whether the type of shock structure that develops in the
simulation depends on the initial gas configuration as well as on the
sound speed. In principle, quasi-stationary gas flows with both types
of shock structure might exist in one and the same gravitational
potential, and the system could evolve towards one or the other
equilibrium, depending on its initial state. To test this, we start a
simulation in an extreme initial state where the system is already in
quasi-equilibrium such that the flow has formed an on-axis shock
structure, and then see whether it evolves to an off-axis shock
solution when we reduce the sound speed to $10\;\KM\SEC^{-1}$.

When the sound speed is dropped, a shock in the original gas flow must
remain a shock.  On the other hand, a shallow compression wave may
suddenly become supersonic, and the newly forming shock can change the
flow pattern in such a way that the original shock structure is
modified or disappears.  When the sound speed is reduced in the
equilibrium flow shown in the top panel of Fig.~\ref{trans}, the main
shock front immediately moves away from the bar major axis in the
direction of the position where the main off-axis shock would be
located in a low-sound speed model. In addition, new spiral shocks
form near the position of the $4/1$-resonance, and much of the gas in
the inner bar region falls towards the centre, building a massive,
elongated $x_1$-disk.  This $x_1$-disk is massive enough to prevent
gas from moving onto $x_2$-orbits, so no off-axis flow can form, and
the whole configuration swings back towards the bar major axis.  The
resulting quasi-equilibrium is shown in the middle panel of
Fig.~\ref{trans}.

We now turn on gas recycling, which essentially removes the high
density material in the $x_1$-disk. To speed up the computation, we
use a higher gas recycling constant
$\alpha=3\;\MSUN^{-1}\;\PC^2\;\GYR^{-1}$ than in the other models.
This removes the obstacle in the flow, and after a few more rotations
the gas now settles on $x_2$-orbits and an off-axis shock front has
formed.  The structure of the inner $x_2$-disk in the final
configuration (bottom panel of Fig.~\ref{trans}) is somewhat different
from those shown previously because of the large gas recycling rate.

We draw two conclusions from this experiment. (i) If we release small
quantities of gas in a barred potential, then independent of the
initial velocity field the flow pattern adjusts itself to that which
is natural for gas at this sound speed in this gravitational
potential.  For, the velocity field of the gas that evolved towards the
final off-axis configuration in Fig.~\ref{trans} must have been close
to the velocity field typical for on-axis flows. (ii) On the other
hand, a massive disk set up previously is hard to perturb and may
prevent the incoming gas flow from taking up its natural flow pattern.
In real galactic disks, such massive disks are expected to form stars
and be depleted of gas; thus eventually we expect the gas there to
return to its preferred configuration.

\subsection{Influence of gas recycling and bar mass}

\begin{figure*}
\epsfxsize=0.91\textwidth\ifpsfiles\epsfbox{fig7.ps}\fi
\caption[-]{%
Influence of bar mass, sound speed and gas recycling on the gas flow
pattern. Models in the left-hand panels have sound speed $\csnd=15\,\KMS$,
those in the right-hand panels $\csnd=20\,\KMS$. Upper two panels show gas
flows in the standard potential and without gas recycling that mimicks
star formation, the middle two panels show gas flows in the standard
potential with gas recycling, and the lower two panels are for a
potential with a bar twice as massive as before and again include gas
recycling.  See Table~1 for parameters. There is no influence of gas
recycling on the flow pattern, but a stronger bar or a higher sound
speed can drive the gas flow towards a configuration with on-axis
shocks.
}
\label{param-seq}
\end{figure*}

In Fig.~\ref{param-seq} we compare models with different sound speeds,
bar masses, and with or without gas recycling as indicated in the
figure caption and Table~1.

First we consider the influence of the gas recycling by comparing
Model \runBD\ with \runBF\ and Model \runBA\ with \runBG.  In both cases
the large scale flow configuration does not change. However the
simulations with gas recycling (\runBF\ \& \runBG) look more clumpy;
we believe this is an artifact of having to introduce gas
particles with different masses in the gas recycling algorithm.

Between the middle panels and the lower panels of Fig.~\ref{param-seq}
the bar mass is doubled and therefore the orbits change as well as the
resonances. The more massive bar forces more elongated $x_1$-orbits
(see Fig.~\ref{flow-A} below).  For the lower sound speed the pattern
maintains off-axis shocks, but the shock front now starts nearer to
the end of the bar and is closer to the bar major axis. There is also
a new large gap in the shock front at $(x,y)=(2,2) \KPC$.

For the higher sound speed (Model \runAC) the flow field in the more
massive bar potential has changed to an on-axis configuration.  Thus
the critical sound speed which divides the two shock patterns appears
to decrease for increasing bar mass.

\subsection{Comparison of SPH and grid based results}

In Model \runcmp\ we have reproduced model~001 of Athanassoula (1992)
to test our code against a grid based method; in this case, a second
order flux splitting code written by G.~D.~van Albada 
(van Albada \& Roberts 1981).  Both methods
solve the Euler equations, thus the same results should be expected.
But differences may arise due to the statistical nature of SPH or the
different numerical and artificial viscosities used in both codes.
Comparing a simulation of the same physical problem computed with
both numerical methods may give valuable clues about their reliability. 

Fig.~\ref{flow-A} shows the particle distribution, gas velocity field,
and closed orbit structure for Model \runcmp. The potential is the
same as used for model~001 of Athanassoula (1992), and the initial
conditions are set up identically except that the disk is truncated
at a smaller radius ($6\,\KPC$ as opposed to $16\,\KPC$ in the grid
model). Compared to the simulations discussed earlier, we have used
a larger smoothing parameter ($\zeta=1.8$). The reason
for this is that, because of the stronger bar, the gas in this
potential moves to the center more quickly, leaving fewer particles
in the shock front in quasi-equilibrium. To measure the shock properties
displayed in Fig.~\ref{shock1-A}, it was then necessary to employ
more smoothing. 

The global structure of the flow seen in Fig.~\ref{flow-A} is similar
to that of the grid model (see Athanassoula's Fig.~2 for
comparison). The point on the major axis where the $4/1$-spiral shock
begins is at the same distance from the centre (at $\simeq 4.3 \KPC$).
The cusped orbit shocks are straight in both simulations, however, in
the SPH model they form a smaller angle with the bar major axis than
in the grid model.
 
The shock properties shown in Fig.~\ref{shock1-A} are similar to
Athanassoulas findings (her Fig.~11) when we take into account that
our and her slits are not at identical positions and that the
resolution is different. The shape of the function describing the gas
velocity along the slit is very similar, and the maximum upstream
value and the downstream velocities agree quantitatively (to be able
to make the comparison with Athanassoula, the velocities in the two
lower panels of Fig.~\ref{shock1-A} are measured relative to the
inertial frame and include a contribution from the bar rotation). The
shock is slightly less well resolved in the SPH simulation,
corresponding to a somewhat broader density peak. The peak density is
underestimated in both simulations; in the SPH case it is also
time-dependent. The value we measure from Fig.~\ref{shock1-A} appears
to be higher than that in Athanassoula's Fig.~11 (the units in both
diagrams are $\Sigma_0 = 1 \MSUN/\PC^2$).

The large scatter in the density plot in Fig.~\ref{shock1-A} is due to
the rather large width of the slit, $4 h$. Particularly in the low
density region (upstream, inside the cusped orbit) there are very few
particles in the SPH simulation. This is a general
resolution problem with SPH: particles follow the gas flow; thus low
density regions near resonances, through which the flow moves
rapidly, are often not well-resolved, whereas regions of high density
such as the inner $x_2$-ring contain many more particles than
necessary for adequate resolution. The reader should note that this
is even true in a symmetrized 2D simulation with 20000 particles.
In three-dimensional simulations, especially when high 
density contrasts are present, resolution may be a much more severe 
problem.

However, as the comparison shows, SPH gives surprisingly
accurate solutions in the case at hand. Thus we may confidently
move to fully self-gravitating simulations in which the real
strengths of SPH will be used.

\newlength{\size}
\setlength{\size}{0.30\textheight}
\begin{figure}
\epsfxsize=0.95\colwidth\ifpsfiles\epsfbox{fig8.ps}\fi
\caption[-]{%
Same as Fig.~\ref{flow-S} for Model \runcmp\ at $t=0.4\;\GYR$.
This model is the same as Model 001 of Athanassoula (1992), and
the figure should be compared with her Fig.~2.
}
\label{flow-A}
\end{figure}
\begin{figure}
\epsfxsize=\colwidth \ifpsfiles\epsfbox{fig9.ps}\fi
\caption[-]{%
Same as Fig.~\ref{shock1-S} for the Slit in Fig.~\ref{flow-A}.
}
\label{shock1-A}
\end{figure}

\section{Discussion}

We have studied the structure of quasi-stationary shock fronts in
realistic barred galaxy models with an ILR, and have found that both
off-axis and on-axis shocks can result in the same rotating potential,
depending on the sound speed of the isothermal fluid. The off-axis
shocks at low sound speeds had previously been assumed to be the rule
in such potentials, whereas on-axis shocks had been found in
potentials too shallow to have an ILR. We have verified in one case
that the gas flow resulting from our two-dimensional SPH simulation
agrees approximately with that found in a grid-based simulation of
Athanassoula (1992). 

If the quasi-equlibrium flow that results in a given galaxy model
depends on the fluid parameters in the hydrodynamical 
calculation --- here, the sound speed --- then it may equally depend
on the entire prescription of modelling the ISM: whether its
complicated multi-phase structure is better represented as particle-like
or in terms of an ideal fluid. This offers the opportunity of
learning about the `correct' description of interstellar fluids from
detailed studies of galactic flows.

The critical sound speed that we have found to divide the two shock
regimes is around $\sim 20\KMS$ in our standard model potential, which
corresponds to a barred galaxy with a circular velocity of $v_c\simeq
200\KMS$. It is somewhat lower for a model with a stronger bar. A
dependence on bar strength etc.\ is not surprising since the
occurrence of a shock depends on both the sound speed and the fluid
flow velocities in the relevant bar region.

Expressed dimensionlessly, the critical sound speed is $\sim 10\%$
of the circular velocity. This implies that dwarf or Magellanic
galaxies should predominantly be in the on-axis regime. This might
have important implications for the dependence of morphology on absolute
magnitude.

For our Galaxy, the value of $\sim 20\KMS$ is in an interesting
regime, for while the cloud velocity dispersion in the Galactic disk
inside the solar radius is $\sim 5 \KMS$ (Clemens 1985), the vertical
cloud velocity dispersion in the Galactic bulge region is inferred to
be $\sim 25\KMS$ (Bally \etal 1988).  Moreover, the effective sound
speed in the ISM may be higher than the cloud dispersion if magnetic
fields contribute significant pressure, as may be expected especially
near the Galactic Centre (Morris 1994).

Assuming that our Galaxy is representative, it would thus appear
difficult to predict the type of flow pattern which would form in a
barred galaxy. It is even unclear whether the flow would find a single
quasi-stationary flow configuration.  There also arises the
possibility that large-scale star formation, by increasing the
turbulent pressure in the disk, might itself change the structure of
the flow by which it was initiated. Since the on-axis shock patterns
are generally associated with larger mass inflow rates, a starburst
changing the pattern from off-axis to on-axis could thereby prolong
its own gas supply. One may even speculate that at early times, when
star formation rates are generally higher, the morphology of the gas
flows could have been systematically shifted towards on-axis shock
flows with higher mass accretion rates. If true this would have
obvious relevance to HST observations of high-redshift galaxies; thus
further work in this direction should be useful.

\section*{Acknowledgments}

We are greatful to M.~Steinmetz for many discussions on SPH and for
making his code available. We also thank E.~Athanassoula, J.~Binney
and M.~Samland for helpful discussions and suggestions. This work was
supported by the Max-Planck Institut f\"ur Astrophysik in Garching
and the Landessternwarte Heidelberg.


\begin{thebibliography}{99}

\bibitem[Athanassoula 1992]{athana92} 
	Athanassoula E., 1992, MNRAS, 259, 345
\bibitem[Balsara 1995]{Balsara95}
	Balsara D.S., 1995, J. Comp. Phys. 121, 357
\bibitem[Bally et al. 1988]{bally88}
	Bally J., Stark A.A., Wilson R.W., Henkel C., 1988, ApJ, 324, 223
\bibitem[Benz 1990]{benz90}
	Benz W., 1990, in Buchler J.~R., ed, 
        The Numerical Modelling of Nonlinear Stellar Pulsations,
	Kluwer, Dordrecht, p. 269
\bibitem[Binney et al. 1991]{BGSBU}    
        Binney J.J., Gerhard O.E., Stark A.A., Bally J., Uchida K.I.,
        1991, MNRAS, 252, 210
\bibitem[Binney \& Gerhard 1993]{BinGer93}
	Binney J.J., Gerhard O.E., 1993, in Back to the Galaxy, AIP
        Conf.~Proc.~278, Holt S.S., Verter F., eds, AIP, New York, p. 87 
\bibitem[Clemens 1985]{Cle85}
	Clemens D.P., 1985, ApJ, 295, 422
\bibitem[Combes \& Gerin 1985]{ComGer85}
	Combes F., Gerin M., 1985, A\&A, 150, 327
\bibitem[Contopoulos \& Papayannopoulos 1980]{conto80}
	Contopoulos G., Papayannopoulos T., 1980, A\&A, 92, 33
\bibitem[Cowie 1980]{Cowie80}  
	Cowie L.~L., 1980, ApJ, 236, 868
\bibitem[Duval \& Athanassoula 1983]{DuvAth83}
	Duval M.F., Athanassoula E., 1983, A\&A, 121, 297
\bibitem[Friedli \& Benz 1993]{FriBen93}
	Friedli D., Benz W., 1993, A\&A, 268, 65
\bibitem[Gerhard \& Binney 1993]{GerBin93}
	Gerhard O.E., Binney J.J., 1993, in Galactic Bulges, Proc.\ IAU
	Symp.~153, Dejonghe H., Habing H.J., eds, Kluwer, Dordrecht, 275
\bibitem[Habe \& Ikeuchi 1985]{HabIke85}
	Habe A., Ikeuchi S., 1985, ApJ, 289, 540
\bibitem[Hernquist \& Katz 1989]{HerKat89}
	Hernquist L., Katz N., 1989, ApJS, 97, 231
\bibitem[Ishizuki et al. 1990]{Ishizuki90}
	Ishizuki S., Kawabe R., Ishiguro M., Okumura S.K., Morita, K.-I., 
	1990, Nat, 344, 224
\bibitem[Jenkins \& Binney 1994]{JenBin94}
	Jenkins A., Binney J.J., 1994, MNRAS, 270, 703
\bibitem[Kenney et al. 1992]{Kenney92}
	Kenney J.D.P., Wilson C.D., Scoville N.Z., Devereux N.A., 
	Young J.S., 1992, ApJ, 395, L79
\bibitem[Lindblad \& J\"ors\"ater 1987]{LinJoer87}
	Lindblad P.O., J\"ors\"ater S., 1988, in Palous J., ed.,
	Evolution of Galaxies. Czech. Acad. Sciences, p.~289
\bibitem[Liszt \& Burton 1980]{LisBur80}
	Liszt H.S., Burton W.B., 1980, ApJ, 236, 779
\bibitem[Monaghan 1992]{Mona92}   
	Monaghan  J.~J., 1992, ARA\&A, 30, 543
\bibitem[Monaghan \& Gingold 1983]{MonGin83}
	Monaghan J.~J., Gingold R.A., 1983, J. Comput. Phys., 52, 374
\bibitem[Morris 1994]{Mor94}
	Morris M., 1994, in Genzel R., Harris A.I., eds, The Nuclei
	of Normal Galaxies. Kluwer, Dordrecht, p.~185
\bibitem[Mulder \& Liem 1986]{MulLie86}
	Mulder W.A., Liem B.T., 1986, A\&A, 157, 148
\bibitem[Pence \& Blackman 1984a]{PenBla1984a}
	Pence W.D., Blackman C.P., 1984a, MNRAS, 207, 9
\bibitem[Pence \& Blackman 1984b]{PenBla1984b}
	Pence W.D., Blackman C.P., 1984b, MNRAS, 210, 547
\bibitem[Quillen \etal 1995]{Quillen95}
	Quillen A.C., Frogel J.A., Kenney J.D.P., Pogge R.W., DePoy D.L.,
        1995, ApJ, 441, 549
\bibitem[Roberts \etal 1979]{Roberts}
	Roberts W.~W., van Albada G.D., Huntley J.M., 1979, ApJ, 233, 67
\bibitem[Sanders \& Huntley 1976]{SanHun76}
	Sanders R.H., Huntley J.M., 1976, ApJ, 209, 53
\bibitem[Sanders \& Tubbs 1980]{SanTub80}
	Sanders R.H., Tubbs A.D., 1980, ApJ, 235, 803
\bibitem[Schmidt 1959]{Schmidt59}
	Schmidt M., 1959, ApJ, 129, 243
\bibitem[Schwarz 1981]{Schwarz81}
	Schwarz M.P., 1981, ApJ, 247, 77
\bibitem[Schwarz 1984]{Schwarz84}
	Schwarz M.P., 1984, MNRAS, 209, 93
\bibitem[Sellwood \& Wilkinson 1993]{SelWil93}
	Sellwood J.A., Wilkinson A., 1993, Rep. Prog. Phys., 56, 173
\bibitem[Steinmetz \& M\"uller 1993]{Steinm93} 
	Steinmetz M., M\"uller E., 1993, A\&A, 268, 391
\bibitem[Syer \& Narayan 1993]{SyNa93}
	Syer D., Narayan R., 1993, MNRAS, 262, 749
\bibitem[Teuben \etal]{Teuben86}
	Teuben P.J., Sanders R.H., Atherton P.D., van Albada G.D.,
	1986, MNRAS, 221, 1
\bibitem[van Albada 1985]{vAlbada85}
	van Albada G.D., 1985, A\&A, 142, 491
\bibitem[van Albada \& Roberts W.~W. 1981]{AlbSan81}
	van Albada T.~S., Roberts W.~W., 1981, ApJ, 246, 740
\bibitem[van Albada \& Sanders 1982]{AlbSan82}
	van Albada T.~S., Sanders R.~H., 1982, MNRAS, 201, 303
\bibitem[Weiner, Williams \& Sellwood 1993]{Weiner93}
	Weiner B.J., Williams T.B., Sellwood J.A., 1993, 
	BAAS, 183, \#76.07

\end{thebibliography}
\end{document}